\documentclass[prb,twocolumn,showpacs,preprintnumbers,amsmath,amssymb]{revtex4-1}
\usepackage{graphicx}
\usepackage{dcolumn}
\usepackage{bm}

\begin{document}

\title{Three Dimensional Magnetism and Coupling to
the Conduction Electrons in PdCrO$_2$}

\author{Khuong P. Ong}
\affiliation{Institute of High Performance Computing,
1 Fusionopolis Way, Singapore 138632}

\author{David J. Singh}
\affiliation{Materials Science and Technology Division,
Oak Ridge National Laboratory, Oak Ridge, Tennessee 37831-6056}

\date{\today}

\begin{abstract}
We report density functional calculations addressing the electronic
structure and magnetic properties of delafossite PdCrO$_2$.
We find substantial magnetic interactions in the $c$-axis direction as
well as beyond first nearest neighbors in-plane, so that PdCrO$_2$
is a 3D frustrated antiferromagnet. We also find substantial coupling
between the Cr moments and the Pd derived conduction electrons.
\end{abstract}

\pacs{75.10.Lp,75.30.Et}

\maketitle

\section{introduction}

Recent discoveries pointing to novel ground states and
unusual behaviors in field and temperature
dependent magnetic and thermodynamic
properties have led to renewed interest
in the physics of frustrated spin systems.
Delafossite structure PdCrO$_2$ has attracted recent attention
in this context.
Importantly, like the sister
compound PdCoO$_2$, it can be grown in high quality form facilitating
experimental investigation of its physical properties.
\cite{takatsu,takatsu2,takatsu3}
This material can be described as a stacking of 2D CrO$_2$ layers,
separated by Pd, with the Cr ions arranged on a triangular lattice.
In a standard
nearest neighbor Heisenberg picture, the ground state would be
expected to be a non-collinear 120$^\circ$ spin structure below
the Neel temperature, $T_N$=37.5 K,
\cite{mekata}
but the actual magnetic structure of PdCrO$_2$ could be more complex.
\cite{rastelli}
Takatsu and co-workers find that the magnetic Bragg peaks in
neutron scattering are rather broad
and half-integer $l$ (1/3,1/3,$l$) and (2/3,2/3,$l$)
peaks are found in addition to the integer $l$ peaks. \cite{takatsu}
In any case, PdCrO$_2$ is a particularly interesting system because
like its sister compound PdCoO$_2$ it is conducting,
\cite{wichainchai}
and has a very large conductivity anisotropy.
\cite{takatsu10}
This offers
an opportunity to study the interplay between frustrated magnetism
and charge carriers in a near two dimensional metal.
Furthermore, recent experiments have revealed an unconventional
anomalous Hall effect in this material.
\cite{takatsu2}

Doumerc and co-workers reported susceptibility data up to $\sim$ 450 K.
\cite{doumerc}
They obtained an effective moment
of 4.1 $\mu_B$ and a Weiss temperature of -500 K,
in a Curie-Weiss fit,
although they qualify these values
as approximate. The data show a poor fit except over a relatively
small temperature range above $\sim$ 300 K, in contrast
to the other compounds they report: CuCrO$_2$, AgCrO$_2$ and CuFeO$_2$,
which show very good Curie-Weiss behavior over a wide temperature
range extending to below 100 K.
In fact, in their data,
there is still an apparent upward curvature to $\chi^{-1}(T)$
at the highest $T$, suggesting that the Weiss temperature could be
lower in magnitude or that the system may have itinerant character.

Takatsu and co-workers show more details of the susceptibility below
350 K.
\cite{takatsu}
There is no plausible Curie-Weiss fit to the data, but they do
state that their data are consistent with the values
extracted by Doumerc and co-workers. Interestingly, the
data shows a broad maximum above $T_N$. However, in contrast to
strongly 2D magnetic systems, $\chi$ just above $T_N$ is only
slightly lower than the maximum value, and there are strong signatures
of the ordering in $\chi(T)$ as well as resistivity and specific heat.
This is not really consistent with expectations for a strongly 2D
Heisenberg system. On the other hand, Takatsu and co-workers do observe
a sublinear resistivity, $\rho(T)$ above $T_N$, which they attribute to
short range magnetic correlations, and an unusual behavior in
the specific heat also above $T_N$.

Motivated by this we performed density functional
calculations of electronic
structure and energetics of PdCrO$_2$
in order to address the nature and role of the metallic conduction
electrons in PdCrO$_2$ and the extent to which the system
is a realization of a 2D nearest neighbor Heisenberg magnet.
We find in contrast to previous
assumptions that
(1) there is a considerable interplay between the Cr moments
and the metallic electrons,
and
(2) there are strong magnetic interactions along the $c$-axis
direction, so that even though PdCrO$_2$ is very two dimensional as 
an electron gas, it is a very three dimensional magnetic system.
Finally, there is metal-metal bonding, consistent with previous
studies of delafossite compounds. \cite{seshadri}

\section{approach}

We performed density functional calculations
with the Perdew, Burke, Ernzerhof (PBE)
generalized gradient approximation, \cite{pbe}
using the
all electron linearized augmented planewave (LAPW) method
\cite{singh-book} as implemented in the WIEN2k code, \cite{wien}
similar to our previous calculations for PdCoO$_2$.
\cite{ong}
We carefully tested the convergence of our results against the
various parameters, including tests with different sphere radii
and different choices of the augmentation
including both LAPW and so-called APW+lo basis sets.
\cite{sjostedt}
We found the results
to be stable.
We used the experimental lattice parameters, $a$=2.923 \AA{}
and $c$=18.087 \AA{} (delafossite structure, space group $R\bar{3}m$,
Pd at ($\frac{1}{2}$,$\frac{1}{2}$,$\frac{1}{2}$),
Cr at (0,0,0), O at ($\pm z$,$\pm z$,$\pm z$),
in rhombohedral
coordinates). \cite{shannon}
Refinement of the
internal parameter corresponding to the O height above the
Cr plane has not been reported in literature to our knowledge.
As such, we calculated this parameter for a ferromagnetic
ordering and used it in the other calculations.
We obtained $z$=0.6111, which yields a Cr-O distance
of 1.962 \AA. This is in accord with the expected bond length
for high spin Cr$^{3+}$ (the Shannon radii are 1.22 \AA{} for O$^{2-}$
and 0.755 \AA{} for octahedral Cr$^{4+}$, summing to 1.975 \AA).
We also note that the Pd-Pd distance of $a$=2.923 \AA{} is
not much longer than the Pd-Pd distance of 2.75 \AA{} in Pd metal.
The Cr-O-Cr bond angles in the layers are 96.3$^\circ$.
Considering the Cr lattice, the O atoms directly connect
the nearest neighbor Cr atoms in plane. There are no direct O connections
to second neighbors. Secondly, in the out-of-plane direction,
the shortest hopping paths are Cr-O-Pd-O-Cr.
The Cr-O-Pd bond angles along this path are 120.7$^\circ$ with this O
position.

\section{moment formation and electronic structure}

We find that PdCrO$_2$ is
strongly unstable against Cr moment formation regardless of 
the arrangement of the moments with an energy of $\sim$1 eV/Cr.
We also find all arrangements are metallic in accord with experiment.
Therefore, PdCrO$_2$ should be described as
a metal containing Cr local moments.
We start our description with the calculated
electronic structure with ferromagnetic
alignment of these moments.
The calculated energy is 0.92 eV lower than the non-spin-polarized
(no Cr moments) case, and the magnetization is 2.87 $\mu_B$ per formula
unit. The magnetization inside a Cr LAPW sphere (radius 2.05 bohr)
is 2.57 $\mu_B$.

\begin{figure}
\includegraphics*[height=0.95\columnwidth,angle=270]{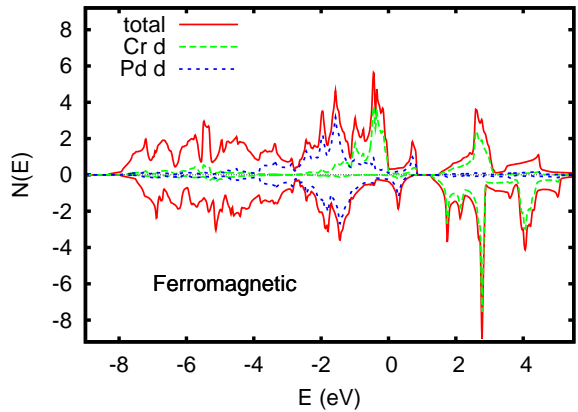}
\caption{(color online) Density of states and projections
on Cr and Pd LAPW spheres of radius 2.05 bohr for ferromagnetic ordering}
\label{dos-f}
\end{figure}

\begin{figure}
\includegraphics*[width=0.45\columnwidth,angle=0]{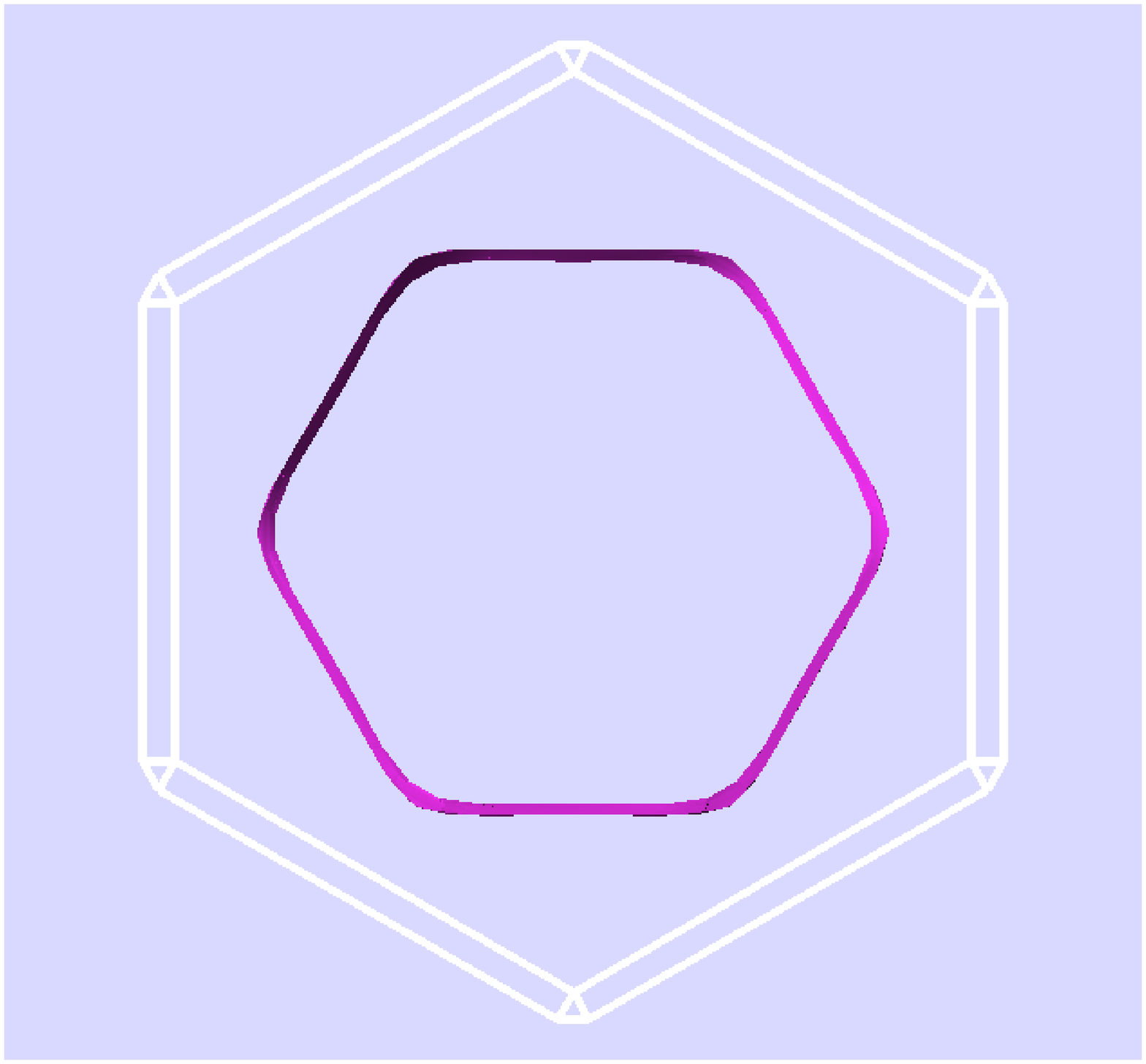}
\includegraphics*[width=0.45\columnwidth,angle=0]{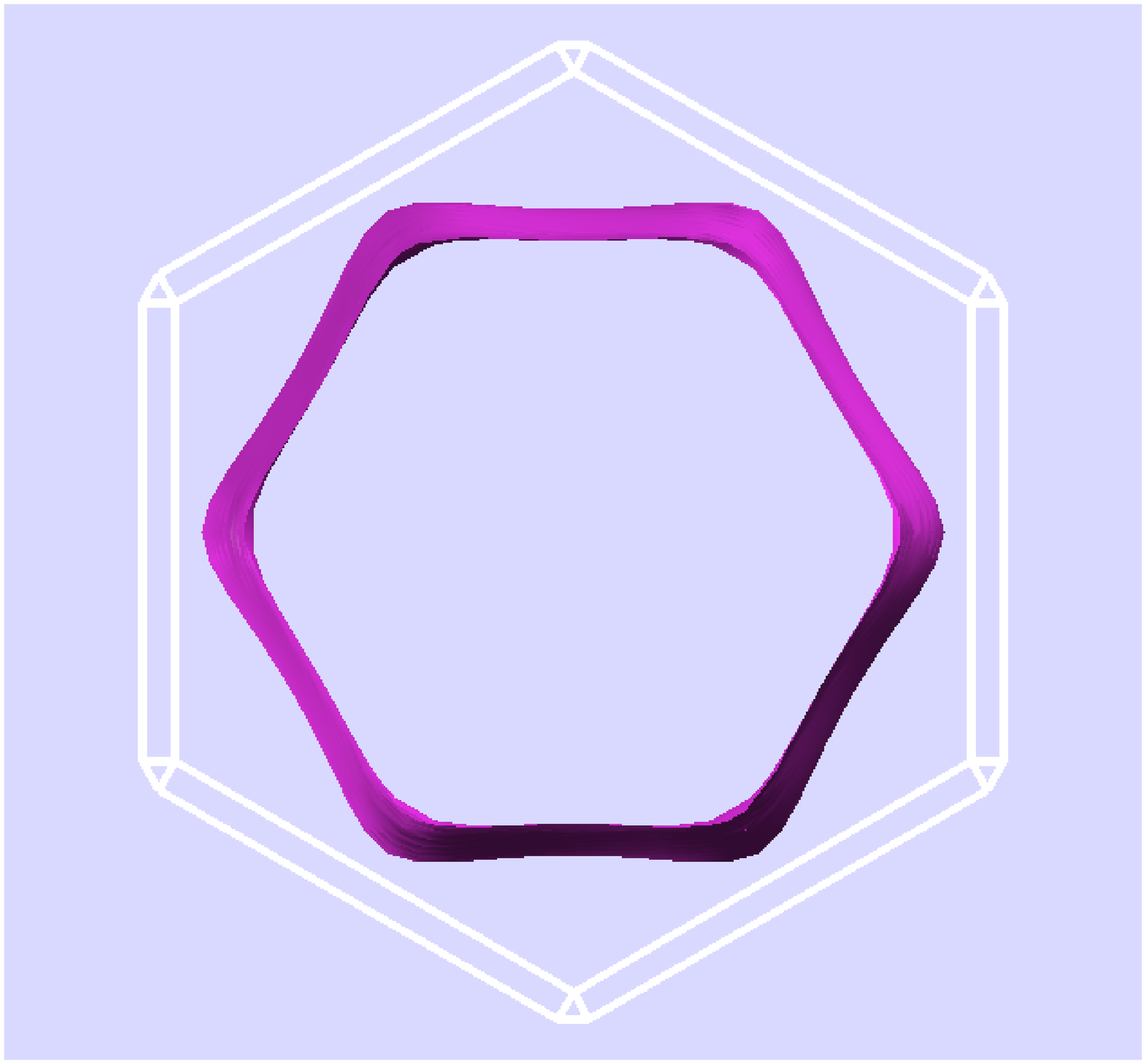}
\caption{(color online) Fermi surface for ferromagnetic ordering.
Left panel is majority spin, right panel is minority. Note
that these are hole cylinders centered at $\Gamma$.}
\label{fs-f}
\end{figure}

The calculated electronic density of states (DOS) is given
in Fig. \ref{dos-f}. It clearly shows crystal field split
Cr $d$ states with the majority spin $t_{2g}$ states occupied, and
the $e_g$ unoccupied, as are all the minority spin Cr states. The
DOS at $E_F$ is Pd derived. These states provide a simple very two dimensional
Fermi surface
(Fig. \ref{fs-f}), consistent with the observed strong conductivity
anisotropy and
reminiscent of PdCoO$_2$, which is also a highly anisotropic metal.
\cite{ong}
As discussed previously, the states giving rise to the Fermi surface
arise from Pd-Pd metal-metal bonding as in other delafossites. \cite{seshadi}
Additionally, two additional features are seen:
(1) The Fermi surface has very little rounding, which is the condition
for maximizing nesting (but note that
density of states is Pd derived and low, $N(E_F)$=0.7 eV$^{-1}$,
per formula unit on a both spins basis
for this ferromagnetic ordering), and
(2) While both the majority and minority spin Fermi surfaces
are similar in shape, they are not identical, implying an interplay
between the metallic, largely Pd derived, conduction band, and the
Cr derived magnetic moments. In fact, the minority spin Fermi
surface is 13\% larger in volume than the majority spin surface.
This interplay between Cr moments and the conduction electrons
provides a mechanism for spin scattering, that will increase the
resistivity as $T$ disorders the spins and also
a mechanism for long range Cr-Cr interactions in-plane, as well as
interactions in the $c$-axis direction.

\section{magnetic interactions}

\begin{figure}
\includegraphics*[width=\columnwidth,angle=0]{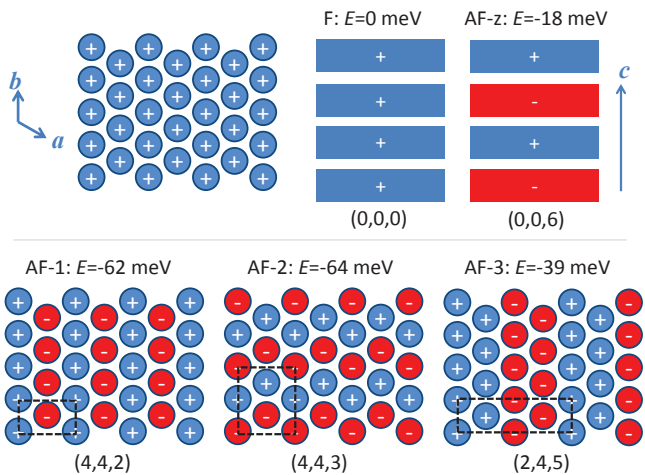}
\caption{(color online) Structures and energetics
of small supercells,
in meV per Cr, relative to the ferromagnetic ordered configuration.
The top panel is for ferromagnetic
layers, stacked along the $c$-axis either ferromagnetically, or
antiferromagnetically, while the bottom panel shows three in-plane
antiferromagnetic orderings.
The unit cells are indicated by dotted black lines.
The three numbers in parentheses below each structure
denote the numbers of opposite spin neighbors,
at the the nearest neighbor position (out of six),
the in-plane next nearest Cr neighbor position (out of six) and
the out-of-plane nearest neighbor position (out of six), respectively.}
\label{hex}
\end{figure}

We studied the magnetic interactions through a series of
supercell calculations with different magnetic orderings.
These calculations and their energetics are summarized in
Fig. \ref{hex}.
One may immediately note from the top panel of the figure
that the out-of-plane magnetic interactions are large.
Specifically, an antiferromagnetic stacking of ferromagnetic
layers is found to be 18 meV/Cr lower in energy than a ferromagnetic
stacking. This is large compared with the ordering temperature,
$T_N$= 38 K, i.e. $kT_N$= 3.3 meV.

\begin{figure}
\includegraphics[width=0.80\columnwidth,angle=0]{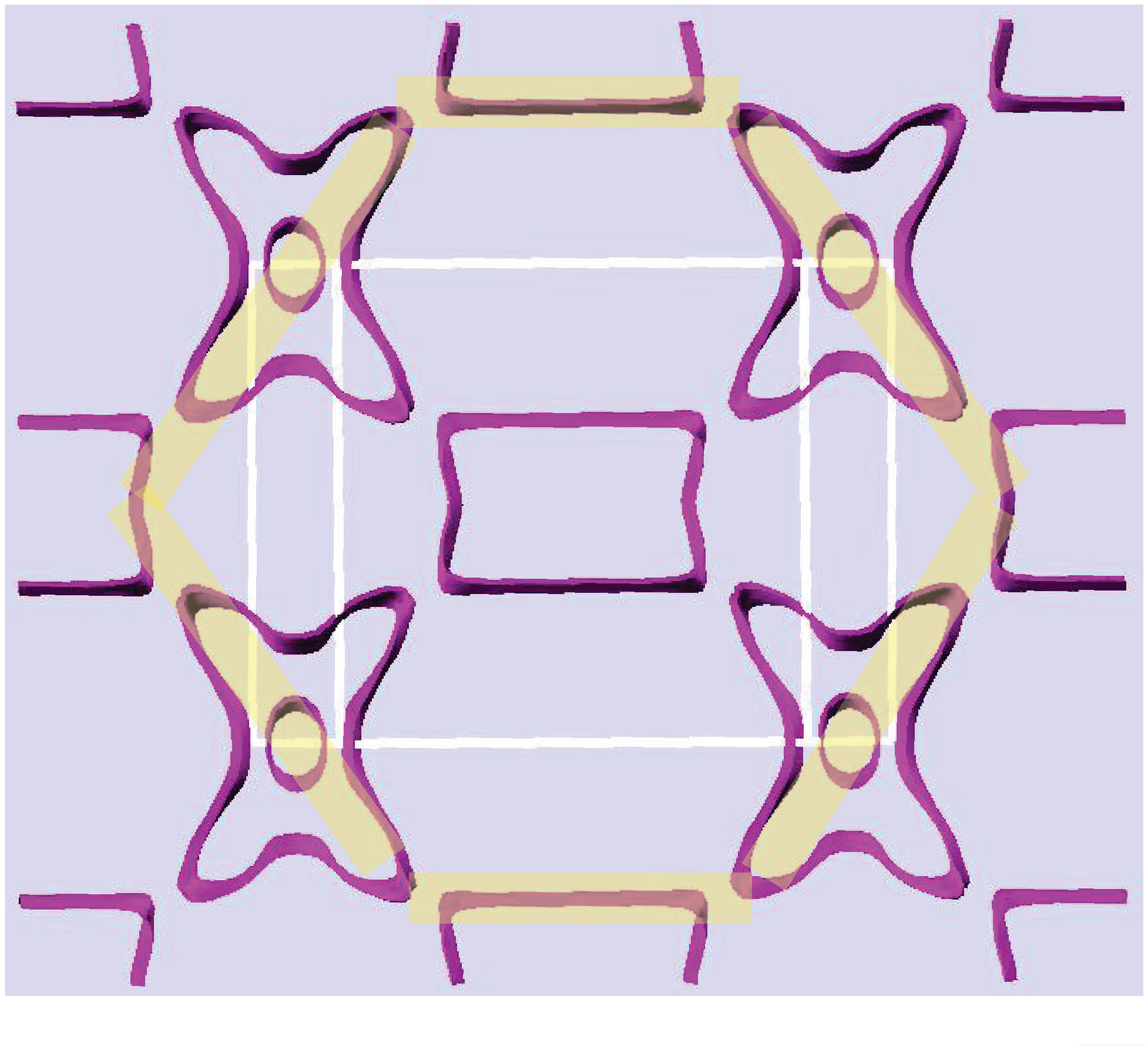}
\caption{(color online) Fermi surface
viewed down the $c$-axis for the lowest energy
AF-2 configuration (see text), shown in an extended zone scheme.
The white lines show the supercell Brillouin zone and the 
highlight sketches the non-reconstructed Fermi surfaces
of Fig. \ref{fs-f}.}
\label{fs-af}
\end{figure}

Of the small collinear cells that we studied, the lowest
energy belongs to AF-2 (see Fig. \ref{hex} for notation),
which corresponds to the ordering in CuFeO$_2$.
\cite{terada,ye,nakajima}
The electronic structure for the various the orderings remains very
anisotropic at the Fermi energy, where
the states are Pd derived. However, as mentioned above,
there is a strong interplay between the Fermi surface and the Cr
moments.
For example, the Fermi surface for the lowest energy magnetic configuration
is shown in Fig. \ref{fs-af}. The figure shows overlapped hexagonal
sections, as expected for the folded zone, but also sizable splittings
at the intersections.

This interplay between magnetic order and the conduction electrons is also 
seen in the calculated densities of states.
As mentioned, for the ferromagnetic order, we obtain $N(E_F)$=0.70 eV$^{-1}$
on a per formula unit both spins basis. For the AF-z (see Fig. \ref{hex})
order, we obtain $N(E_F)$=0.69 eV$^{-1}$ on the
same basis, while we obtain
$N(E_F)$= 0.81 eV$^{-1}$, for AF-1,
$N(E_F)$= 0.76 eV$^{-1}$, for AF-2, and
$N(E_F)$= 0.77 eV$^{-1}$, for AF-3.
The bare specific heat coefficient $\gamma$ inferred from these
values is 1.6-1.9 mJ/mol K$^2$, comparable to the specific heat
value of Takatsu and co-workers of 1.4$\pm$0.2 mJ/mol K$^{2}$
in the ground state.
\cite{takatsu}
This leaves little room for any renormalization at low $T$.

The variability of the
electronic structure at $E_F$ depending on magnetic state
provides a qualitative framework for understanding why the
resistivity of high quality crystals of PdCrO$_2$ is so much
higher than that of PdCoO$_2$, even though the Fermi surface,
structure and band character at the Fermi surface are very similar,
\cite{ong,ong2}
and also why the resistivity shows a strong signature of the magnetic
ordering, with a pronounced decrease in resistance as $T$ is lowered
through $T_N$.
\cite{takatsu}
Specifically, this connection between magnetic order
and the electrons at the Fermi surface indicates that strong spin-fluctuation
scattering is expected above the ordering temperature, freezing out as
$T$ is lowered below $T_N$.

Materials with this type of coupling between
magnetism and electrons at the Fermi surface tend to be the ones that
display unusual properties, such as superconductivity (as in e.g. the
Fe-pnictides where
the coupling is very strong) when the magnetism is suppressed. Therefore,
this coupling of spin fluctuations to electrons at the Fermi surface suggests
that it will be of considerable interest to
experimentally examine what happens as the
magnetic order is suppressed, e.g. by pressure or alloying.

We now turn to the energetics (Fig. \ref{hex}) in more detail.
To do this we write a short range model with interactions
to nearest and next-nearest neighbors in-plane and to the
nearest out-of-plane neighbor:

\begin{equation}
E= E_{(F)} + j_1 N_{1\downarrow}
+ j_2 N_{2\downarrow}
+ j_z N_{z\downarrow}
\end{equation}

\noindent
where $E$ is the energy per Cr, $E_{(F)}$ is the
energy of the ferromagnetic ordering,
$N_{1\downarrow}$ is the average number of opposite spin
first nearest neighbors in the structure, and similarly for
next nearest neighbors and neighbors along the $c$-axis direction.
While this could be converted into a three neighbor Heisenberg
model, we write it this way because the moments in the Cr
LAPW spheres vary
between the different configurations (in the range 2.48 $\mu_B$ to
2.57 $\mu_B$).

In any case, if we match the energies of the ordered structures,
F, AF-z, AF-1 and AF-3, we obtain
$j_1$= -16 meV, $j_2$= +2 meV and $j_z$= -3 meV.
The energies cannot be reasonably reproduced without all three 
parameters (note that each shell has 6 atoms, so e.g. the difference
between ferromagnetic and AF-z is 6$j_z$=-18 meV, which is not small, etc.)
Use of these values, yields the energy for the AF-2 state to within
1 meV / Cr atom.
While the model is perhaps too simple to capture all aspects of the
magnetic interactions in PdCrO$_2$, it does clearly show that
the material cannot be reasonably understood as a simple 2D
nearest neighbor Heisenberg model.

\section{discussion and conclusions}

Considering the ordering temperature $kT_N$=3.3 meV,
it would seem that none of these interactions can be neglected,
i.e. at least first and second in-plane neighbors and crucially c-axis
interactions need to be included when discussing the magnetism
of PdCrO$_2$.

This conclusion may seem unexpected considering the very two dimensional
electronic structure at the Fermi energy, as found both in our
calculations and as is clear from the experimental resistivity anisotropy.
\cite{takatsu10}
However, we note that a similar behavior is found
in the layered cobaltate Na$_x$CoO$_2$.
\cite{bayrakci,helme}
That materials has a related crystal structure, that is similar
to the delafossite,
except that the layer stacking differs and bridges between planes
are O neighbors along $c$ instead of two O separated by a Pd
(i.e. in Na$_x$CoO$_2$ the O in neighboring layers are
directly on top of each other).
The three dimensional magnetic behavior in Na$_x$CoO$_2$
arises in part because of this bonding topology,
which provides many superexchange
exchange paths between the neighboring CdI$_2$ structure CoO$_2$ layers.
\cite{johannes}
Within this framework, the c-axis magnetic interactions and the
coupling to the conduction electrons are inter-related since they are
both mediated by or through Pd.

In any case, we note that the
Kosterlitz-Thouless type of suppression of the ordering temperature
is logarithmic in the anisotropy of the interactions.
As such, very strong anisotropy is needed
to obtain two dimensional, as opposed to three dimensional, magnetism.
We can conclude that PdCrO$_2$ is a three dimensional frustrated
antiferromagnet with an interesting interplay between magnetism
and conduction electrons.

\acknowledgements

Work at ORNL was supported by the Department of Energy, Basic
Energy Sciences, Materials Sciences and Engineering Division.
Work at IHPC was supported by the Singapore Agency for Science
Technology and Research (A*STAR). DJS is grateful for the
hospitality of IHPC where a portion of this work was performed.

\bibliography{PdCrO2}

\begin{thebibliography}{10}%
\makeatletter
\providecommand \@ifxundefined [1]{%
 \ifx #1\undefined \expandafter \@firstoftwo
 \else \expandafter \@secondoftwo
\fi
}%
\providecommand \@ifnum [1]{%
 \ifnum #1\expandafter \@firstoftwo
 \else \expandafter \@secondoftwo
\fi
}%
\providecommand \enquote [1]{``#1''}%
\providecommand \bibnamefont  [1]{#1}%
\providecommand \bibfnamefont [1]{#1}%
\providecommand \citenamefont [1]{#1}%
\providecommand\href[0]{\@sanitize\@href}%
\providecommand\@href[1]{\endgroup\@@startlink{#1}\endgroup\@@href}%
\providecommand\@@href[1]{#1\@@endlink}%
\providecommand \@sanitize [0]{\begingroup\catcode`\&12\catcode`\#12\relax}%
\@ifxundefined \pdfoutput {\@firstoftwo}{%
 \@ifnum{\z@=\pdfoutput}{\@firstoftwo}{\@secondoftwo}%
}{%
 \providecommand\@@startlink[1]{\leavevmode\special{html:<a href="#1">}}%
 \providecommand\@@endlink[0]{\special{html:</a>}}%
}{%
 \providecommand\@@startlink[1]{%
  \leavevmode
  \pdfstartlink
   attr{/Border[0 0 1 ]/H/I/C[0 1 1]}%
   user{/Subtype/Link/A<</Type/Action/S/URI/URI(#1)>>}%
  \relax
 }%
 \providecommand\@@endlink[0]{\pdfendlink}%
}%
\providecommand \url  [0]{\begingroup\@sanitize \@url }%
\providecommand \@url [1]{\endgroup\@href {#1}{\urlprefix}}%
\providecommand \urlprefix [0]{URL }%
\providecommand \Eprint[0]{\href }%
\@ifxundefined \urlstyle {%
  \providecommand \doi [1]{doi:\discretionary{}{}{}#1}%
}{%
  \providecommand \doi [0]{doi:\discretionary{}{}{}\begingroup
  \urlstyle{rm}\Url }%
}%
\providecommand \doibase [0]{http://dx.doi.org/}%
\providecommand \Doi[1]{\href{\doibase#1}}%
\providecommand \bibAnnote [3]{%
  \BibitemShut{#1}%
  \begin{quotation}\noindent
    \textsc{Key:}\ #2\\\textsc{Annotation:}\ #3%
  \end{quotation}%
}%
\providecommand \bibAnnoteFile [2]{%
  \IfFileExists{#2}{\bibAnnote {#1} {#2} {\input{#2}}}{}%
}%
\providecommand \typeout [0]{\immediate \write \m@ne }%
\providecommand \selectlanguage [0]{\@gobble}%
\providecommand \bibinfo [0]{\@secondoftwo}%
\providecommand \bibfield [0]{\@secondoftwo}%
\providecommand \translation [1]{[#1]}%
\providecommand \BibitemOpen[0]{}%
\providecommand \bibitemStop [0]{}%
\providecommand \bibitemNoStop [0]{.\EOS\space}%
\providecommand \EOS [0]{\spacefactor3000\relax}%
\providecommand \BibitemShut [1]{\csname bibitem#1\endcsname}%
\bibitem{takatsu}%
  \BibitemOpen
  \bibfield{author}{%
  \bibinfo {author} {\bibfnamefont{H.}~\bibnamefont{Takatsu}}, \bibinfo
  {author} {\bibfnamefont{H.}~\bibnamefont{Yoshizawa}}, \bibinfo {author}
  {\bibfnamefont{S.}~\bibnamefont{Yonezawa}},\ and\ \bibinfo {author}
  {\bibfnamefont{Y.}~\bibnamefont{Maeno}},\ }%
  \bibfield{journal}{%
  \bibinfo {journal} {Phys. Rev. B}\ }%
  \textbf{\bibinfo {volume} {79}},\ \bibinfo {pages} {104424} (\bibinfo {year}
  {2009})%
  \bibAnnoteFile{NoStop}{takatsu}%
\bibitem{takatsu2}%
  \BibitemOpen
  \bibfield{author}{%
  \bibinfo {author} {\bibfnamefont{H.}~\bibnamefont{Takatsu}}, \bibinfo
  {author} {\bibfnamefont{S.}~\bibnamefont{Yonezawa}}, \bibinfo {author}
  {\bibfnamefont{S.}~\bibnamefont{Fujimoto}},\ and\ \bibinfo {author}
  {\bibfnamefont{Y.}~\bibnamefont{Maeno}},\ }%
  \bibfield{journal}{%
  \bibinfo {journal} {Phys. Rev. Lett.}\ }%
  \textbf{\bibinfo {volume} {105}},\ \bibinfo {pages} {137201} (\bibinfo {year}
  {2010})%
  \bibAnnoteFile{NoStop}{takatsu2}%
\bibitem{takatsu3}%
  \BibitemOpen
  \bibfield{author}{%
  \bibinfo {author} {\bibfnamefont{H.}~\bibnamefont{Takatsu}}\ and\ \bibinfo
  {author} {\bibfnamefont{Y.}~\bibnamefont{Maeno}},\ }%
  \bibfield{journal}{%
  \bibinfo {journal} {J. Cryst. Growth}\ }%
  \textbf{\bibinfo {volume} {312}},\ \bibinfo {pages} {3461} (\bibinfo {year}
  {2010})%
  \bibAnnoteFile{NoStop}{takatsu3}%
\bibitem{mekata}%
  \BibitemOpen
  \bibfield{author}{%
  \bibinfo {author} {\bibfnamefont{M.}~\bibnamefont{Mekata}}, \bibinfo {author}
  {\bibfnamefont{T.}~\bibnamefont{Sugino}}, \bibinfo {author}
  {\bibfnamefont{A.}~\bibnamefont{Oohara}}, \bibinfo {author}
  {\bibfnamefont{Y.}~\bibnamefont{Oohara}},\ and\ \bibinfo {author}
  {\bibfnamefont{H.}~\bibnamefont{Yoshizawa}},\ }%
  \bibfield{journal}{%
  \bibinfo {journal} {Physica B}\ }%
  \textbf{\bibinfo {volume} {213}},\ \bibinfo {pages} {221} (\bibinfo {year}
  {1995})%
  \bibAnnoteFile{NoStop}{mekata}%
\bibitem{rastelli}%
  \BibitemOpen
  \bibfield{author}{%
  \bibinfo {author} {\bibfnamefont{E.}~\bibnamefont{Rastelli}}\ and\ \bibinfo
  {author} {\bibfnamefont{A.}~\bibnamefont{Tassi}},\ }%
  \bibfield{journal}{%
  \bibinfo {journal} {J. Appl. Phys.}\ }%
  \textbf{\bibinfo {volume} {81}},\ \bibinfo {pages} {4143} (\bibinfo {year}
  {1997})%
  \bibAnnoteFile{NoStop}{rastelli}%
\bibitem{wichainchai}%
  \BibitemOpen
  \bibfield{author}{%
  \bibinfo {author} {\bibfnamefont{A.}~\bibnamefont{Wichainchai}}, \bibinfo
  {author} {\bibfnamefont{P.}~\bibnamefont{Dordor}}, \bibinfo {author}
  {\bibfnamefont{J.~P.}\ \bibnamefont{Doumerc}}, \bibinfo {author}
  {\bibfnamefont{E.}~\bibnamefont{Marquestaut}}, \bibinfo {author}
  {\bibfnamefont{M.}~\bibnamefont{Pouchard}}, \bibinfo {author}
  {\bibfnamefont{P.}~\bibnamefont{Hagenmuller}},\ and\ \bibinfo {author}
  {\bibfnamefont{A.}~\bibnamefont{Ammar}},\ }%
  \bibfield{journal}{%
  \bibinfo {journal} {J. Solid State Chem.}\ }%
  \textbf{\bibinfo {volume} {74}},\ \bibinfo {pages} {126} (\bibinfo {year}
  {1988})%
  \bibAnnoteFile{NoStop}{wichainchai}%
\bibitem{takatsu10}%
  \BibitemOpen
  \bibfield{author}{%
  \bibinfo {author} {\bibfnamefont{H.}~\bibnamefont{Takatsu}}, \bibinfo
  {author} {\bibfnamefont{S.}~\bibnamefont{Yonezawa}}, \bibinfo {author}
  {\bibfnamefont{C.}~\bibnamefont{Michioka}}, \bibinfo {author}
  {\bibfnamefont{K.}~\bibnamefont{Yoshimora}},\ and\ \bibinfo {author}
  {\bibfnamefont{Y.}~\bibnamefont{Maeno}},\ }%
  \bibfield{journal}{%
  \bibinfo {journal} {J. Phys. Conf. Ser.}\ }%
  \textbf{\bibinfo {volume} {200}},\ \bibinfo {pages} {012198} (\bibinfo {year}
  {2010})%
  \bibAnnoteFile{NoStop}{takatsu10}%
\bibitem{doumerc}%
  \BibitemOpen
  \bibfield{author}{%
  \bibinfo {author} {\bibfnamefont{J.~P.}\ \bibnamefont{Doumerc}}, \bibinfo
  {author} {\bibfnamefont{A.}~\bibnamefont{Wichainchai}}, \bibinfo {author}
  {\bibfnamefont{A.}~\bibnamefont{Ammar}}, \bibinfo {author}
  {\bibfnamefont{M.}~\bibnamefont{Pouchard}},\ and\ \bibinfo {author}
  {\bibfnamefont{P.}~\bibnamefont{Hagenmuller}},\ }%
  \bibfield{journal}{%
  \bibinfo {journal} {Mat. Res. Bull.}\ }%
  \textbf{\bibinfo {volume} {21}},\ \bibinfo {pages} {745} (\bibinfo {year}
  {1986})%
  \bibAnnoteFile{NoStop}{doumerc}%
\bibitem{seshadri}%
  \BibitemOpen
  \bibfield{author}{%
  \bibinfo {author} {\bibfnamefont{R.}~\bibnamefont{Seshadri}}, \bibinfo
  {author} {\bibfnamefont{C.}~\bibnamefont{Felser}}, \bibinfo {author}
  {\bibfnamefont{K.}~\bibnamefont{Thieme}},\ and\ \bibinfo {author}
  {\bibfnamefont{W.}~\bibnamefont{Tremel}},\ }%
  \bibfield{journal}{%
  \bibinfo {journal} {Chem. Mater.}\ }%
  \textbf{\bibinfo {volume} {10}},\ \bibinfo {pages} {2189} (\bibinfo {year}
  {1998})%
  \bibAnnoteFile{NoStop}{seshadri}%
\bibitem{pbe}%
  \BibitemOpen
  \bibfield{author}{%
  \bibinfo {author} {\bibfnamefont{J.~P.}\ \bibnamefont{Perdew}}, \bibinfo
  {author} {\bibfnamefont{K.}~\bibnamefont{Burke}},\ and\ \bibinfo {author}
  {\bibfnamefont{M.}~\bibnamefont{Ernzerhof}},\ }%
  \bibfield{journal}{%
  \bibinfo {journal} {Phys. Rev. Lett.}\ }%
  \textbf{\bibinfo {volume} {77}},\ \bibinfo {pages} {3865} (\bibinfo {year}
  {1996})%
  \bibAnnoteFile{NoStop}{pbe}%
\bibitem{singh-book}%
  \BibitemOpen
  \bibfield{author}{%
  \bibinfo {author} {\bibfnamefont{D.~J.}\ \bibnamefont{Singh}}\ and\ \bibinfo
  {author} {\bibfnamefont{L.}~\bibnamefont{Nordstrom}},\ }%
  \emph{\bibinfo {title} {{Planewaves, Pseudopotentials and the LAPW Method,
  2nd Ed.}}}\ (\bibinfo {publisher} {Springer Verlag},\ \bibinfo {year}
  {2006})%
  \bibAnnoteFile{NoStop}{singh-book}%
\bibitem{wien}%
  \BibitemOpen
  \bibfield{author}{%
  \bibinfo {author} {\bibfnamefont{P.}~\bibnamefont{Blaha}}, \bibinfo {author}
  {\bibfnamefont{K.}~\bibnamefont{Schwarz}}, \bibinfo {author}
  {\bibfnamefont{G.}~\bibnamefont{Madsen}}, \bibinfo {author}
  {\bibfnamefont{D.}~\bibnamefont{Kvasnicka}},\ and\ \bibinfo {author}
  {\bibfnamefont{J.}~\bibnamefont{Luitz}},\ }%
  \bibfield{journal}{%
  \bibinfo {journal} {WIEN2k, An Augmented Plane Wave + Local Orbitals Program
  for Calculating Crystal Properties (K. Schwarz, Tech. Univ. Wien, Austria)}}%
   (\bibinfo {year} {2001})%
  \bibAnnoteFile{NoStop}{wien}%
\bibitem{ong}%
  \BibitemOpen
  \bibfield{author}{%
  \bibinfo {author} {\bibfnamefont{K.~P.}\ \bibnamefont{Ong}}, \bibinfo
  {author} {\bibfnamefont{D.~J.}\ \bibnamefont{Singh}},\ and\ \bibinfo {author}
  {\bibfnamefont{P.}~\bibnamefont{Wu}},\ }%
  \bibfield{journal}{%
  \bibinfo {journal} {Phys. Rev. Lett.}\ }%
  \textbf{\bibinfo {volume} {104}},\ \bibinfo {pages} {176601} (\bibinfo {year}
  {2010})%
  \bibAnnoteFile{NoStop}{ong}%
\bibitem{sjostedt}%
  \BibitemOpen
  \bibfield{author}{%
  \bibinfo {author} {\bibfnamefont{E.}~\bibnamefont{Sjostedt}}, \bibinfo
  {author} {\bibfnamefont{L.}~\bibnamefont{Nordstrom}},\ and\ \bibinfo {author}
  {\bibfnamefont{D.~J.}\ \bibnamefont{Singh}},\ }%
  \bibfield{journal}{%
  \bibinfo {journal} {Solid State Commun.}\ }%
  \textbf{\bibinfo {volume} {114}},\ \bibinfo {pages} {15} (\bibinfo {year}
  {2000})%
  \bibAnnoteFile{NoStop}{sjostedt}%
\bibitem{shannon}%
  \BibitemOpen
  \bibfield{author}{%
  \bibinfo {author} {\bibfnamefont{R.~D.}\ \bibnamefont{Shannon}}, \bibinfo
  {author} {\bibfnamefont{D.~B.}\ \bibnamefont{Rogers}},\ and\ \bibinfo
  {author} {\bibfnamefont{C.~T.}\ \bibnamefont{Prewitt}},\ }%
  \bibfield{journal}{%
  \bibinfo {journal} {Inorg. Chem.}\ }%
  \textbf{\bibinfo {volume} {10}},\ \bibinfo {pages} {713} (\bibinfo {year}
  {1971})%
  \bibAnnoteFile{NoStop}{shannon}%
\bibitem{terada}%
  \BibitemOpen
  \bibfield{author}{%
  \bibinfo {author} {\bibfnamefont{N.}~\bibnamefont{Terada}}, \bibinfo {author}
  {\bibfnamefont{S.}~\bibnamefont{Mitsuda}}, \bibinfo {author}
  {\bibfnamefont{H.}~\bibnamefont{Oshumi}},\ and\ \bibinfo {author}
  {\bibfnamefont{K.}~\bibnamefont{Tajima}},\ }%
  \bibfield{journal}{%
  \bibinfo {journal} {J. Phys. Soc. Jpn.}\ }%
  \textbf{\bibinfo {volume} {75}},\ \bibinfo {pages} {023602} (\bibinfo {year}
  {2006})%
  \bibAnnoteFile{NoStop}{terada}%
\bibitem{ye}%
  \BibitemOpen
  \bibfield{author}{%
  \bibinfo {author} {\bibfnamefont{F.}~\bibnamefont{Ye}}, \bibinfo {author}
  {\bibfnamefont{Y.}~\bibnamefont{Ren}}, \bibinfo {author}
  {\bibfnamefont{Q.}~\bibnamefont{Huang}}, \bibinfo {author}
  {\bibfnamefont{J.~A.}\ \bibnamefont{{Fernandez-Baca}}}, \bibinfo {author}
  {\bibfnamefont{P.}~\bibnamefont{Dai}}, \bibinfo {author}
  {\bibfnamefont{J.~W.}\ \bibnamefont{Lynn}},\ and\ \bibinfo {author}
  {\bibfnamefont{T.}~\bibnamefont{Kimura}},\ }%
  \bibfield{journal}{%
  \bibinfo {journal} {Phys. Rev. B}\ }%
  \textbf{\bibinfo {volume} {73}},\ \bibinfo {pages} {220404(R)} (\bibinfo
  {year} {2006})%
  \bibAnnoteFile{NoStop}{ye}%
\bibitem{nakajima}%
  \BibitemOpen
  \bibfield{author}{%
  \bibinfo {author} {\bibfnamefont{T.}~\bibnamefont{Nakajima}}, \bibinfo
  {author} {\bibfnamefont{A.}~\bibnamefont{Suno}}, \bibinfo {author}
  {\bibfnamefont{S.}~\bibnamefont{Mitsuda}}, \bibinfo {author}
  {\bibfnamefont{N.}~\bibnamefont{Terada}}, \bibinfo {author}
  {\bibfnamefont{S.}~\bibnamefont{Kimura}}, \bibinfo {author}
  {\bibfnamefont{K.}~\bibnamefont{Kaneko}},\ and\ \bibinfo {author}
  {\bibfnamefont{H.}~\bibnamefont{Yamauchi}},\ }%
  \bibfield{journal}{%
  \bibinfo {journal} {Phys. Rev. B}\ }%
  \textbf{\bibinfo {volume} {84}},\ \bibinfo {pages} {184401} (\bibinfo {year}
  {2011})%
  \bibAnnoteFile{NoStop}{nakajima}%
\bibitem{ong2}%
  \BibitemOpen
  \bibfield{author}{%
  \bibinfo {author} {\bibfnamefont{K.~P.}\ \bibnamefont{Ong}}, \bibinfo
  {author} {\bibfnamefont{J.}~\bibnamefont{Zhang}}, \bibinfo {author}
  {\bibfnamefont{J.~S.}\ \bibnamefont{Tse}},\ and\ \bibinfo {author}
  {\bibfnamefont{P.}~\bibnamefont{Wu}},\ }%
  \bibfield{journal}{%
  \bibinfo {journal} {Phys. Rev. B}\ }%
  \textbf{\bibinfo {volume} {81}},\ \bibinfo {pages} {115120} (\bibinfo {year}
  {2010})%
  \bibAnnoteFile{NoStop}{ong2}%
\bibitem{bayrakci}%
  \BibitemOpen
  \bibfield{author}{%
  \bibinfo {author} {\bibfnamefont{S.~P.}\ \bibnamefont{Bayrakci}}, \bibinfo
  {author} {\bibfnamefont{I.}~\bibnamefont{Mirebeau}}, \bibinfo {author}
  {\bibfnamefont{P.}~\bibnamefont{Bourges}}, \bibinfo {author}
  {\bibfnamefont{Y.}~\bibnamefont{Sidis}}, \bibinfo {author}
  {\bibfnamefont{M.}~\bibnamefont{Enderle}}, \bibinfo {author}
  {\bibfnamefont{J.}~\bibnamefont{Mesot}}, \bibinfo {author}
  {\bibfnamefont{D.~P.}\ \bibnamefont{Chen}}, \bibinfo {author}
  {\bibfnamefont{C.~T.}\ \bibnamefont{Lin}},\ and\ \bibinfo {author}
  {\bibfnamefont{B.}~\bibnamefont{Keimer}},\ }%
  \bibfield{journal}{%
  \bibinfo {journal} {Phys. Rev. Lett.}\ }%
  \textbf{\bibinfo {volume} {94}},\ \bibinfo {pages} {157205} (\bibinfo {year}
  {2005})%
  \bibAnnoteFile{NoStop}{bayrakci}%
\bibitem{helme}%
  \BibitemOpen
  \bibfield{author}{%
  \bibinfo {author} {\bibfnamefont{L.~M.}\ \bibnamefont{Helme}}, \bibinfo
  {author} {\bibfnamefont{A.~T.}\ \bibnamefont{Boothroyd}}, \bibinfo {author}
  {\bibfnamefont{R.}~\bibnamefont{Coldea}}, \bibinfo {author}
  {\bibfnamefont{D.}~\bibnamefont{Prabhakaran}}, \bibinfo {author}
  {\bibfnamefont{A.}~\bibnamefont{Stunault}}, \bibinfo {author}
  {\bibfnamefont{G.~J.}\ \bibnamefont{{McIntyre}}},\ and\ \bibinfo {author}
  {\bibfnamefont{N.}~\bibnamefont{Kernavanois}},\ }%
  \bibfield{journal}{%
  \bibinfo {journal} {Phys. Rev. B}\ }%
  \textbf{\bibinfo {volume} {73}},\ \bibinfo {pages} {054405} (\bibinfo {year}
  {2006})%
  \bibAnnoteFile{NoStop}{helme}%
\bibitem{johannes}%
  \BibitemOpen
  \bibfield{author}{%
  \bibinfo {author} {\bibfnamefont{M.~D.}\ \bibnamefont{Johannes}}, \bibinfo
  {author} {\bibfnamefont{I.~I.}\ \bibnamefont{Mazin}},\ and\ \bibinfo {author}
  {\bibfnamefont{D.~J.}\ \bibnamefont{Singh}},\ }%
  \bibfield{journal}{%
  \bibinfo {journal} {Phys. Rev. B}\ }%
  \textbf{\bibinfo {volume} {71}},\ \bibinfo {pages} {214410} (\bibinfo {year}
  {2005})%
  \bibAnnoteFile{NoStop}{johannes}%
\end{thebibliography}%

\end{document}